\documentclass[reprint, showkeys, 
amsmath,amssymb, aps, prl,
]{revtex4-2}

\usepackage{graphicx}
\usepackage{dcolumn}
\usepackage{bm}
\usepackage{hyperref}

\begin{document}

\preprint{test1234}

\title{Solving formally the  Auxiliary System of $O(N)$ Non Linear Sigma Model}

\author{Dimitrios Katsinis}
 \altaffiliation[Also at ]{Department of Physics, National and Kapodistrian University of Athens, University Campus, 15784, Zografou, Athens, Greece.}
  \altaffiliation[Also at ]{NCSR “Demokritos”, Institute of Nuclear and Particle Physics, 15310, Agia Paraskevi, Attiki, Greece.}
 \email{dkatsinis@phys.uoa.gr}
\affiliation{Instituto de F\'isica, Universidade de S\~ao Paulo, Rua do Mat\~ao Travessa 1371, 05508-090 S\~ao Paulo, SP, Brazil}

%
%

\date{\today}

\begin{abstract}
We show that the integrability of the $SO(N)/SO(N-1)$ Principal Chiral Model (PCM) originates from the Pohlmeyer reduction of the $O(N)$ Non Linear Sigma Model (NLSM). In particular, we show that the Lax pair of the PCM is related upon redefinitions and identification of parameters to the zero curvature condition, which is a consequence of the flatness of the enhanced space used in the Pohlmeyer reduction. This identification provides the solution of the auxiliary system that corresponds to an \emph{arbitrary} NLSM/PCM solution.
\end{abstract}

\maketitle


\section{Introduction}
\label{sec:intro}

The classical integrability of the PCM is guaranteed by the formulation of its equations of motion and the zero curvature condition, which is satisfied automatically by the current, as a Lax pair \cite{Zakharov:1973pp}. For a recent review see \cite{Zarembo:2017muf}. Both equations can be derived as the compatibility condition of the auxiliary system:
\begin{equation}\label{eq:auxiliary_Psi}
\partial_\pm \Psi(\lambda)=\mathcal{L}_\pm\Psi(\lambda),\qquad \mathcal{L}_\pm=\frac{\left(\partial_\pm g\right)g^{-1}}{1\pm\lambda},
\end{equation}
where $\lambda\in\mathbb{C}$ is the spectral parameter. In the literature, this system also goes by the name fundamental linear problem. In our previous work \cite{Katsinis:2020avk} we solved these equations for $g\in SO(3)/SO(2)$. This was achieved in a brute force manner, which was inspired by the application of the dressing method on elliptic strings in $\mathbb{R}\times\textrm{S}^2$ \cite{Katsinis:2018ewd}, as well as on the elliptic  minimal surfaces in $\textrm{H}^3$ \cite{Katsinis:2020dhe}. We introduced spherical coordinates and bootstrapped the solution of the auxiliary system using all the available equations. In this work, we will solve these equations for an arbitrary $g\in SO(N)/SO(N-1)$ in a more elegant way. 

A key element of our derivation will be the so-called Pohlmeyer reduction of the $O(N)$ NLSM \cite{Pohlmeyer:1975nb,Lund:1976ze}. In a nutshell, considering the target space $\textrm{S}^{N-1}$ embedded in $\mathbb{R}^N$, the embedding equations of the NLSM solution in $\mathbb{R}^N$ are multicomponent generalizations of the sine-Gordon equations, known as Symmetric Space Sine Gordon models, for a recent review see \cite{Miramontes:2008wt}. This process is the Pohlmeyer reduction.  It is important to point out that the Pohlmeyer reduction is a many-to-one mapping. In particular, there is an infinite family of NLSM solutions, which correspond to the same Pohlmeyer counterpart. Finally, it worth noticing that for a given solution of the Pohlmeyer reduced theory, the equations of motion of the NLSM become linear.

The solution of the auxiliary system \eqref{eq:auxiliary_Psi} is of great interest since it is related to the monodromy matrix, which provides an infinite tower of conserved  charges. In particular, the monodromy matrix $T(\sigma_f,\sigma_i;\lambda)$ is defined as
\begin{equation}
T(\sigma_f,\sigma_i;\lambda)=\Psi(\tau,\sigma_f;\lambda)\Psi^{-1}(\tau,\sigma_i;\lambda).
\end{equation}
It is straightforward to show that its time derivative reads
\begin{multline}
\partial_\tau T(\sigma_f,\sigma_i;\lambda)= \mathcal{L}_\tau(\tau,\sigma_f)T(\sigma_f,\sigma_i;\lambda)\\-T(\sigma_f,\sigma_i;\lambda)\mathcal{L}_\tau(\tau,\sigma_i).
\end{multline}
The Lax connection $\mathcal{L}_\tau$ appears in the auxiliary system when it is expressed in terms of the world-sheet coordinates $\sigma$ and $\tau$. In particular, the relevant equation reads $\partial_\tau \Psi(\tau,\sigma;\lambda)=\mathcal{L}_\tau\Psi(\tau,\sigma;\lambda)$. The values of $\sigma_i$ and $\sigma_f$ are related to the specific solution and its boundary conditions. The latter also specify the explicit form of the conserved quantity. Let us make this statement clear. If $\mathcal{L}_\tau$ vanished both at $\sigma_i$ and $\sigma_f$, which for instance may happen at $\pm\infty$, then the monodromy matrix is conserved.  On the contrary, when imposing periodic boundary conditions, it is its trace that is conserved. Finally, let us also mention that in the case of open - string boundary conditions one can employ the formalism of Cherdnik and Sklyanin by defining a boundary monodromy matrix \cite{Cherednik:1984vvp,Sklyanin:1988yz}.

There is another use for the solution of the axiliary system. It is the necessary input for the application of the dressing method \cite{Zakharov:1973pp,Zakharov:1980ty,Harnad:1983we}. The latter is a technique which enables the construction of new solutions of the NLSM, once a solution is already known. The known solution is referred to as the seed solution. The Pohlmeyer reduced theory has a counterpart of the dressing transformations, known as the B\"acklund transformations. Interestingly enough, a dressing transformation on the NLSM solution, automatically performs a B\"acklund transformation at the avatar of the solution in the Pohlmeyer reduced theory \cite{Hollowood:2009tw}.

In \cite{Katsinis:2020avk} we claimed that the systematic solution of the auxiliary system corresponding to the $O(3)$ NLSM, points to a non-linear superposition rule for NLSMs on symmetric spaces. In order to obtain a new NLSM solution one can combine a NLSM solution with a ``virtual'' one, i.e. one that obeys inadmissible Virasoro constraints when the model is embedded in string theory, both corresponding to the same Pohlmeyer counterpart. We verify this statement in the case of $O(N)$ NLSM.

\section{The Pohlmeyer Reduction of the $O(N)$ NLSM}
\label{sec:pohl}
In this section we review the Pohlmeyer reduction of the $O(N)$ NLSM. The main reason for doing so is that we will use slightly unusual conventions, which will facilitate the rest of the paper. The equations of motion of the NLSM read
\begin{equation}\label{eq:NLSM_EOM}
\partial_{+}\partial_{-}\vec{X}+\left(\partial_+\vec{X}\cdot\partial_-\vec{X}\right)\vec{X}=0,
\end{equation}
where we used the coordinates $\xi^\pm=\xi^1\pm \xi^0$. They are accompanied by the geometric constraint
\begin{equation}\label{eq:geometric}
\vert\vec{X}\vert^2=1,
\end{equation}
as well as by the conservation of the stress-energy tensor $T$, which in appropriate coordinates implies
\begin{equation}\label{eq:Virasoro}
T_{\pm\pm}=\vert\partial_\pm\vec{X}\vert^2=m^2_\pm.
\end{equation}

We introduce a basis $\mathcal{V}=\{\vec{v}_1,\dots,\vec{v}_{N}\}$ in $\mathbb{R}^N$, where $\vec{v}_{N-2}=\partial_- \vec{X}$, $\vec{v}_{N-1}=\partial_+ \vec{X}$ and $\vec{v}_{N}=\vec{X}$ \footnote{In the following, lower-case English letters will always run from 1 to $N-3$, while lower-case Greek letters will run from 1 to $N$.}. On the rest of the vectors we impose the orthonormality conditions
\begin{equation}
\vec{v}_i\cdot \vec{v}_j=\delta_{ij},\quad \vec{v}_i\cdot\vec{v}_{N-2}=\vec{v}_i\cdot\vec{v}_{N-1}=\vec{v}_i\cdot\vec{v}_{N}=0.
\end{equation}
The norms of the last three vectors are fixed by equations \eqref{eq:geometric} and \eqref{eq:Virasoro}. As a consequence, in the subspace spanned by the vectors $\vec{X}$ and $\partial_\pm\vec{X}$ the only unconstrained degree of freedom is the angle between $\partial_+\vec{X}$ and $\partial_-\vec{X}$. Thus, we define the primary Pohlmeyer field $\varphi$ through
\begin{equation}\label{eq:Pohl_def}
\partial_+\vec{X}\cdot \partial_-\vec{X}=m_+ m_-\cos\varphi.
\end{equation}
Let us expand the derivatives of the vectors, which form the basis, on the basis itself as
\begin{equation}
\partial_\pm \vec{v}_{\alpha} =\left(A_{\pm}\right)_{\alpha\beta} \vec{v}_\beta,
\end{equation}
or in matrix form as
\begin{equation}\label{eq:def_basis_matrix}
\partial_\pm V=A_{\pm}V,\qquad V_{ij}=\left(\vec{v}_i\right)_j.
\end{equation}

By definition $\partial_+\vec{v}_N=\vec{v}_{N-1}$ and $\partial_-\vec{v}_N=\vec{v}_{N-2}$, implying that the elements $\left(A_+\right)_{N\alpha}=\delta_{\alpha(N-1)}$ and $\left(A_-\right)_{N\alpha}=\delta_{\alpha(N-2)}$. The equations of motion imply that $\left(A_+\right)_{(N-2)\alpha}=-m_+ m_-\cos\varphi \delta_{\alpha N}$ and similarly $\left(A_-\right)_{(N-1)\alpha}=-m_+ m_-\cos\varphi \delta_{\alpha N}$. Taking into account the following inner products
\begin{align}
&\left(\partial_\pm^2\vec{X}\right)\cdot\vec{X}=-m^2_\pm,\\
&\left(\partial_\pm^2\vec{X}\right)\cdot\partial_\pm\vec{X}=0,\\
&\left(\partial_\pm^2\vec{X}\right)\cdot\partial_\mp\vec{X}=-m_+ m_-\sin\varphi\,\partial_\pm\varphi,
\end{align}
we obtain
\begin{eqnarray}
\partial_\pm^2\vec{X}=-m^2_{\pm}\vec{X}+\cot\varphi\,\partial_\pm\varphi\,\partial_\pm \vec{X}& \nonumber\\
-\frac{m_\pm}{m_\mp}\frac{\partial_\pm\varphi}{\sin\varphi}\partial_\mp \vec{X} &+ m_\pm a^{\pm}_i \vec{v}_i,
\end{eqnarray}
where $a^{\pm}_i$ are additional Pohlmeyer fields, which are grouped as two $(N-3)\times 1$ column matrices ${\bf a}_{\pm}$ \footnote{In the following, symbols in bold will denote column matrices}. Finally, we define
\begin{equation}
\vec{v}_j\cdot\partial_\pm\vec{v}_i=\left(\mathcal{A}_{\pm}\right)_{ij},
\end{equation}
where $\mathcal{A}_\pm$ are $(N-3)\times(N-3)$ antisymmetric matrices. Thus, the matrices $A^{\pm}$ are given by:
\begin{equation}
A_{+}=\begin{pmatrix}
\mathcal{A}_+ & \frac{\cos\varphi\, {\bf a}_+}{m_-\sin^2\phi} & -\frac{{\bf a}_+}{m_+\sin^2\phi} &{\bf 0} \\
{\bf 0}^T & 0 &0 &{\displaystyle -m_+ m_-\cos\varphi} \\
{\displaystyle m_+ {\bf a}_+^T} & -\frac{m_+}{m_-}\frac{\partial_+\varphi}{\sin\varphi} & {\displaystyle \cot\varphi\,\partial_+\varphi} &-m^2_+\\
{\bf 0}^T &0 & 1&  0
\end{pmatrix}
\end{equation}
and
\begin{equation}
A_{-}=\begin{pmatrix}
\mathcal{A}_- &  -\frac{{\bf a}_-}{m_-\sin^2\phi} & \frac{\cos\varphi\, {\bf a}_-}{m_+\sin^2\phi} &{\bf 0} \\
{\displaystyle m_- {\bf a}_-^T} & {\displaystyle \cot\varphi\,\partial_-\varphi} & -\frac{m_-}{m_+}\frac{\partial_-\varphi}{\sin\varphi}  &-m^2_-\\
{\bf 0}^T & 0 &0 &{\displaystyle -m_+ m_-\cos\varphi} \\
{\bf 0}^T &1 & 0&  0
\end{pmatrix}.
\end{equation}
The compatibility condition $\partial_+\partial_-\vec{v}_i=\partial_-\partial_+\vec{v}_i$, is equivalent to the zero curvature condition
\begin{equation}
\partial_- A_+-\partial_+ A_-+\left[A_+,A_-\right]=0.
\end{equation}
Explicitly, the equations of motion of the Pohlmeyer reduced theory read:
\begin{align}
&F_{+-}=\frac{\cos\varphi}{\sin^2\varphi}\left[{\bf a}_-{\bf a}_+^T-{\bf a}_+{\bf a}_-^T\right]\label{eq:EOM_Pohl_1}\\
&\mathcal{D}_{\pm}{\bf a}_\mp=\frac{\partial_\mp\varphi}{\sin\varphi} {\bf a}_\pm\label{eq:EOM_Pohl_2}\\
&\partial_+\partial_-\varphi+\frac{{\bf a}_+^T {\bf a}_-}{\sin\varphi}+m_+ m_-\sin\varphi=0\label{eq:EOM_Pohl_3},
\end{align}
where $\mathcal{D}_{\pm}=I_{N-3}\partial_\pm-\mathcal{A}_\pm$ and $F_{+-}=\left[\mathcal{D}_+,\mathcal{D}_-\right]$ \footnote{$I_{N-3}$ denotes the $(N-3)\times(N-3)$ identity matrix.}. Notice that the Pohlmeyer reduced theory depends only on the product $m_+m_-$. This is crucial for the rest of the paper. It seems that we ended up facing a more complicated problem than the original NLSM. However, this is not the case since the physical degrees of freedom are actually only $N-2$. Let us define the $(N-2)\times (N-2)$ matrices
\begin{equation}\label{eq:def_A_flat}
\tilde{A}_+ =\begin{pmatrix}
\mathcal{A}_+ & -\frac{{\bf a}_+}{\tan\varphi}\\
\frac{{\bf a}_+^T}{\tan\varphi} &0
\end{pmatrix},\quad
\tilde{A}_-=\begin{pmatrix}
\mathcal{A}_- & \frac{{\bf a}_-}{\sin\varphi}\\
-\frac{{\bf a}_-^T}{\sin\varphi} &0
\end{pmatrix}.
\end{equation}
Notice that $\tilde{A}_\pm^T=-\tilde{A}_\pm$, thus $\tilde{A}_\pm$ are valued in the Lie algebra $\mathfrak{so}(N-2)$. The equations of motion \eqref{eq:EOM_Pohl_1}-\eqref{eq:EOM_Pohl_3} imply that $\tilde{A}_\pm$ constitute a flat connection, i.e.
\begin{equation}\label{eq:ZCC_tilde}
\partial_- \tilde{A}_+-\partial_+ \tilde{A}_-+\left[\tilde{A}_+,\tilde{A}_-\right]=0.
\end{equation}
Moreover, equation \eqref{eq:EOM_Pohl_2} assumes the form
\begin{align}
\tilde{\mathcal{D}}_-{\bf \tilde{a}}_+&=\frac{{\bf \tilde{a}}_-^T{\bf \tilde{a}}_+}{\sin\varphi}{\bf \tilde{Z}}+\frac{\partial_+\varphi}{\sin\varphi} {\bf \tilde{a}}_-,\label{eq:EOM_Pohl_2a}\\
\tilde{\mathcal{D}}_+{\bf \tilde{a}}_-&=-\frac{{\bf \tilde{a}}_-^T{\bf \tilde{a}}_+}{\tan\varphi}{\bf \tilde{Z}}+\frac{\partial_-\varphi}{\sin\varphi} {\bf \tilde{a}}_+,\quad {\bf \tilde{a}}_\pm=\begin{pmatrix}
{\bf a}_\pm\\ 0\label{eq:EOM_Pohl_2b}
\end{pmatrix},
\end{align}
where ${\bf \tilde{Z}}_i=\delta_{i,(N-2)}$ and  $\tilde{\mathcal{D}}_{\pm}=I_{N-2}\partial_\pm-\tilde{\mathcal{A}}_\pm$. Finally, equation \eqref{eq:EOM_Pohl_3} assumes the form
\begin{equation}\label{eq:EOM_Pohl_3_tilde}
\partial_+\partial_-\varphi+\frac{\tilde{\bf a}_+^T \tilde{\bf a}_-}{\sin\varphi}+m_+ m_-\sin\varphi=0
\end{equation}

The solution of equation \eqref{eq:ZCC_tilde} is $\tilde{A}_\pm=\left(\partial_\pm \tilde{O}\right)\tilde{O}^T$, where $\tilde{O}\in SO(N-2)$, which implies that
\begin{align}
{\bf \tilde{a}}_+=\tan\varphi\, \tilde{O}\,\partial_+ {\bf Z},\quad {\bf \tilde{a}}_-=-\sin\varphi\, \tilde{O}\,\partial_- {\bf Z},
\end{align}
where ${\bf Z}=\tilde{O}^T{\bf \tilde{Z}}$ is of unit norm. It is straightforward to show that equations \eqref{eq:EOM_Pohl_2a} and \eqref{eq:EOM_Pohl_2b} are equivalent to
\begin{multline}
\partial_+\partial_- {\bf Z}+\frac{\partial_+\varphi}{\tan\varphi}\partial_-{\bf Z}+\frac{\partial_-\varphi}{\sin\varphi\cos\varphi}\partial_+{\bf Z}\\+\left(\partial_- {\bf Z}^T\partial_+ {\bf Z}\right){\bf Z}=0,
\end{multline}
while equation \eqref{eq:EOM_Pohl_3} becomes
\begin{equation}
\partial_+\partial_-\varphi-\tan\varphi\left(\partial_- {\bf Z}^T\partial_+ {\bf Z}\right)+m_+ m_-\sin\varphi=0.
\end{equation}
These two equations, describing the $N-2$ physical degrees of freedom, appeared in \cite{Pohlmeyer:1979ch}. One can obtain another set of equations by interchanging $\tan\varphi\leftrightarrow -\sin\varphi$ in \eqref{eq:def_A_flat}. Actually, the equations of motion can be derived  from a $SO(N-1)/SO(N-2)$ gauged Wess-Zumino-Witten model perturbed by a potential term implementing an appropriate gauge fixing \cite{Bakas:1995bm}, see also \cite{Miramontes:2008wt}. Remnants of symmetry are also present in the above construction. Had $\tilde{\bf Z}$ been a dynamical field,  the theory described by equations \eqref{eq:ZCC_tilde}-\eqref{eq:EOM_Pohl_3_tilde} would have the following $SO(N-2)$ gauge redundancy 
\begin{multline}
\tilde{\bf a}_\pm\rightarrow \tilde{O} \tilde{\bf a}_\pm,\qquad \tilde{\bf Z}\rightarrow \tilde{O} \tilde{\bf Z},\\
\mathcal{S}\mathcal{A}_\pm\rightarrow \tilde{O}   \tilde{\mathcal{S}}\tilde{\mathcal{A}}_\pm \tilde{O}^T+\left(\partial_\pm \tilde{O}\right) \tilde{O}^T.
\end{multline}
Since $\tilde{\bf Z}$ has a specific value, one may interpret this fact as a gauge fixing condition.
 
\section{The Auxiliary System}
The solution of the NLSM,  henceforth denoted as the column $X$, corresponds to an element $g$ of the coset $SO(N)/SO(N-1)$ through the mapping
\begin{equation}
g=\theta(I-2XX^T),\qquad \theta=(I-2X_0X_0^T),
\end{equation}
where $I$ is the $N\times N$ identity matrix, $X_0$ is a constant vector and both $X$ and $X_0$ are of unit norm. By construction $g$ obeys:
\begin{equation}
\bar{g}=g,\qquad g\theta g\theta=I,\qquad g^T=g^{-1},
\end{equation}
implying that indeed $g\in SO(N)/SO(N-1)$. Let us define $\hat{\Psi}$ via the equation \footnote{This is a different redefinition than the one used in our previous works \cite{Katsinis:2020avk,Katsinis:2018ewd}. The redefinition used in the dressing of elliptic strings in $\mathbb{R}\times\textrm{S}^2$ results in a pair of matrices multiplying $\hat{\Psi}$, which depend on a sole world-sheet coordinate. This selection facilitates the solution of the auxiliary system using the explicit form of the elliptic string string solutions. It turns out that there is a more efficient way to handle the case of arbitrary seed solution.}
\begin{equation}\label{eq:Psi_hat_def}
\Psi=g\hat{\Psi}.
\end{equation}
It is straightforward to show that the auxiliary system assumes the form
\begin{equation}\label{eq:auxiliary_Psi_hat_gf}
\partial_\pm\hat{\Psi}=\pm\frac{2\lambda}{1\pm\lambda}\hat{j}_\pm\hat{\Psi},
\end{equation}
where $\hat{j}_\pm=-\frac{1}{2}g^{-1}\partial_\pm g$. Its explicit form is
\begin{equation}\label{eq:j_tilde}
\hat{j}_\pm=\left(\partial_\pm X\right)X^T-X\partial_\pm X^T.
\end{equation}
For reasons that will become apparent shortly, we let
\begin{equation}\label{eq:auxiliary_Psi_tilde}
\hat{\Psi}=V^{-1}\Delta^{-1}\tilde{\Psi},\quad \Delta=\begin{pmatrix}
I_{N-3} & 0 & 0 &0\\
0 & \frac{1+\lambda}{1-\lambda} &0 &0 \\
0 & 0 &  \frac{1-\lambda}{1+\lambda} &0 \\
0 & 0 & 0 & 1
\end{pmatrix}
\end{equation}
where $V$ is the matrix introduced in equation \eqref{eq:def_basis_matrix}. Taking equation \eqref{eq:auxiliary_Psi_tilde} into account, as well as equation \eqref{eq:def_basis_matrix}, the auxiliary system \eqref{eq:auxiliary_Psi_hat_gf} assumes the form
\begin{equation}\label{eq:tildePsi}
\partial_\pm\tilde{\Psi}=\Delta\left[ A_{\pm}\pm \frac{2\lambda}{1\pm\lambda} V \hat{j}_\pm V^{-1}\right]\Delta^{-1}\tilde{\Psi}.
\end{equation}
Equation \eqref{eq:def_basis_matrix} implies that the matrix $V$ is given by
\begin{equation}\label{eq:V_matrix}
V^T=\begin{pmatrix}
v_1 & \dots & v_{N-2} & v_{N-1} & v_{N}
\end{pmatrix}.
\end{equation}
The inverse of this matrix, i.e. $V^{-1}$ is
\begin{equation}\label{eq:V_inv_matrix}
V^{-1}=\begin{pmatrix}
v_1 & \dots & v_{N-2}^\prime & v_{N-1}^\prime & v_{N}
\end{pmatrix},
\end{equation}
where $v_{N-2}^\prime$ and $v_{N-1}^\prime$ are given by
\begin{align}
\vec{v}_{N-2}^{\,\prime}&=\frac{1}{m_-^2\sin^2\varphi}\left[\partial_-\vec{X}-\frac{m_-}{m_+}\cos\varphi\,\partial_+\vec{X}\right],\\
\vec{v}_{N-1}^{\,\prime}&=\frac{1}{m_+^2\sin^2\varphi}\left[\partial_+\vec{X}-\frac{m_+}{m_-}\cos\varphi\,\partial_-\vec{X}\right].
\end{align}
Using equation \eqref{eq:V_matrix}, \eqref{eq:V_inv_matrix} and \eqref{eq:j_tilde}, one obtains
\begin{align}
V \hat{j}_+ V^{-1}&=\begin{pmatrix}
0_{N-3} & {\bf 0} & {\bf 0} & {\bf 0} \\
{\bf 0}^T & 0 & 0 & m_+ m_-\cos\varphi \\
{\bf 0}^T & 0 & 0 & m^2_+\\
{\bf 0}^T & 0 & -1 &  0
\end{pmatrix}\\
V \hat{j}_- V^{-1}&=\begin{pmatrix}
0_{N-3} & {\bf 0} & {\bf 0} & {\bf 0} \\
{\bf 0}^T & 0 &0 & m^2_- \\
{\bf 0}^T & 0 & 0 & m_+ m_-\cos\varphi\\
{\bf 0}^T & -1 &0 &  0
\end{pmatrix}
\end{align}
where $0_{N-3}$ is the $(N-3)\times(N-3)$ zero matrix. Thus, it is straightforward to show equations \eqref{eq:tildePsi} actually assume the form
\begin{equation}\label{eq:tildePsi_final}
\partial_\pm\tilde{\Psi}=\left[A_{\pm}\vert_{m_{\pm}\rightarrow\frac{1\mp\lambda}{1\pm\lambda}m_\pm}\right]\tilde{\Psi},
\end{equation}
which in view of equation \eqref{eq:def_basis_matrix} implies that $\tilde{\Psi}$ is given by
\begin{equation}\label{eq:Psi_tilde_sol}
\tilde{\Psi}=V\vert_{m_{\pm}\rightarrow\frac{1\mp\lambda}{1\pm\lambda}m_\pm}.
\end{equation}
Putting everything together, $\Psi$ reads
\begin{equation}\label{eq:Psi_final}
\Psi(\lambda)=gV^{-1}\Delta^{-1}\left(V\vert_{m_{\pm}\rightarrow\frac{1\mp\lambda}{1\pm\lambda}m_\pm}\right).
\end{equation}

This calculation indicates that given the whole family of NLSM solutions, which correspond to the same solution of the Pohlmeyer reduced theory, one can trivially construct the solution of the auxiliary system. This is possible because the transformation $m_{\pm}\rightarrow\frac{1\mp\lambda}{1\pm\lambda}m_\pm$ leaves invariant the Pohlmeyer reduced theory. In a sense, the fact that the Pohlmeyer reduction is a many-to-one mapping, generates the spectral parameter $\lambda$.

One could multiply the solution by any constant matrix as $\Psi\rightarrow\Psi C$, we choose the normalization used for the application of the dressing method, namely 
\begin{equation}\label{eq:Psi_0}
\Psi(0)=g.
\end{equation} 
As the elements of the matrix $V$ are real functions of the real parameters $m_\pm$, it is evident that $\Psi(\lambda)$ obeys by construction the reality condition
\begin{equation}
\bar{\Psi}\left(\bar{\lambda}\right)=\Psi(\lambda).
\end{equation}
It is straightforward to show that
\begin{equation}
\Psi^T \Psi=\Psi\Psi^T=I.
\end{equation}
Finally, the consistency of the auxiliary system \eqref{eq:auxiliary_Psi} for $\lambda\rightarrow1/\lambda$ requires that $g \theta \Psi(1/\lambda)\theta$ belong to the set of solutions of the equations, i.e. $g \theta \Psi(1/\lambda)\theta=\Psi(\lambda)M$ for some constant matrix $M$. By an elementary multiplication of matrices, it turns out that
\begin{equation}
\Delta(\lambda)V\theta gV^{-1}\Delta^{-1}(1/\lambda)=\mathcal{I}, \thickspace\mathcal{I}=\begin{pmatrix}
I_{N-3} & 0 \\
0 & -I_3
\end{pmatrix}.
\end{equation}
Thus, using equations \eqref{eq:Psi_hat_def} and \eqref{eq:auxiliary_Psi_tilde} it follows that $\tilde{\Psi}$ is subject to the constraint
\begin{equation}
\mathcal{I}\tilde{\Psi}(1/\lambda)\theta=\tilde{\Psi}(\lambda)M.
\end{equation}
Finally, taking equation \eqref{eq:Psi_tilde_sol} into account, this constraint is equivalent to
\begin{equation}
V^{-1}\mathcal{I}\left(V\vert_{m_{\pm}\rightarrow-m_{\pm}}\right)\theta=M,
\end{equation}
which has to be satisfied by the vector of the basis. 
\section{Discussion}
In this work we obtained the formal solution of the auxiliary system corresponding to $O(N)$ NLSM. As long as the whole family of NLSM solutions, which correspond to a given solution of the Pohlmeyer reduced theory, is known, the solution of the auxiliary system can be constructed systematically. 

There are various implications of this construction. To begin with, one can calculate the values of the conserved charges directly, by expanding the monodromy matrix. Also, in view of our result, the dressing method is the implementation of a non-linear superposition of solutions with ``virtual'' ones. There are no differential equations to be solved. The same is also true for the B\"acklund transformation of the Pohlmeyer reduced theory. Essentially, the non-linear superposition is the NLSM counterpart of the insertion of solitons in the Pohlmeyer reduced theory. In this spirit, our results, combined with the addition formula for the on-shell action derived in \cite{Katsinis:2020dhe} enable the calculation of instanton contibutions over any classical configuration of the $O(N)$ NLSMs. It would be compelling to use this formalism in order to make contact with studies on the possible path integration contours for such models \cite{Krichever:2020tgp} or to discuss semi-classical quantization.

Regarding the monodromy matrix, the form of the solution of the auxiliary system \eqref{eq:Psi_final} implies that in the case of periodic boundary conditions its eigenvalues coincide with the eigenvalues of the matrix
\begin{equation}\label{eq:monodromy_eigs}
\mathcal{T}=V(\tau,\sigma_f)V^{-1}(\tau,\sigma_i)\vert_{m_\pm\rightarrow\frac{1\mp\lambda}{1\pm\lambda}m_\pm}.
\end{equation}
It is very important to point out that the NLSM solution and correspondingly the coset element $g$ and the matrix $V$ satisfy periodic boundary conditions for specific $m_{\pm}$. The rescaling of these parameters spoils the periodic boundary conditions (at least for arbitrary $\lambda$). This is the mechanism behind the non-trivial monodromy matrix. One should perceive the notation of \eqref{eq:monodromy_eigs} as rescaling $m_{\pm}$ first and then substituting the value of $\sigma$. More details and a worked-out example can be found in the follow up publication \cite{Katsinis:2022hyn}.

Of course, there are intriguing generalizations of these results. For instance, on could study other symmetric spaces or introduce supersymmetry. Besides the mathematical curiosity, such generalizations are also interesting for practical reasons. In the context of AdS/CFT correspondence \cite{Maldacena:1997re,Gubser:1998bc,Witten:1998qj} it is known that classical free IIB superstring theory on $\textrm{AdS}_5\times\textrm{S}^5$ and the operators of non-perturbative planar $\mathcal{N}=4$ Super-Yang- Mills share a spectral curve \cite{Beisert:2005bm}. Generalizing the presented construction for the Metsaev-Tseytlin action \cite{Metsaev:1998it}, it could be the case that string configuration can be related to specific dual operators. Of course this would require the analogous construction on the field theory side. Nevertheless, the presented construction concerns the $SO(6)$ sector in the vector representation, whose spectral curve was presented in \cite{Beisert:2004ag}.

Finally, this construction may be relevant for the study of the stability of classical strings propagating on spheres in the spirit of \cite{Katsinis:2019oox,Katsinis:2019sdo}. Classical strings are unstable whenever superluminal solitons can propagate on the background of the Pohlmeyer counterpart of these configurations. As the boundary conditions are crucial for such studies, it is unsure how feasible it is to study the stability of arbitrary string configurations, yet one could study specific string solutions if available.

\paragraph*{Acknowledgements}
\begin{acknowledgments}
The work of the author was supported by FAPESP grant 2021/01819-0. The author would like to thank G. Pastras for valuable discussions and comments on the manuscript.  Finally, the author is grateful to the organizers of the Young Researchers Symposium, which preceded the International Congruence on Mathematical Physics 2021, for giving him the opportunity to present parts of this work.
\end{acknowledgments}


%

\end{document}